\documentstyle[12pt,citesort,subeqn]{article}

\begin{document}

\title{\bf Surface correlations for two-dimensional Coulomb fluids in a
disc}
\maketitle

\begin{center}
\large
B. Jancovici
\end{center}

\noindent Laboratoire de Physique Th\'eorique, B\^atiment 210,
Universit\'e de Paris-Sud, 91405 \linebreak Orsay, France (Unit\'e Mixte de 
Recherche no.8627-CNRS); e-mail: Bernard.Jancovici\linebreak @th.u-psud.fr

\begin{abstract}
After a brief review of previous work, two exactly solvable
two-dimensional models of a finite Coulomb fluid in a disc are
studied. The charge correlation function near the boundary circle is
computed. When the disc radius is large compared to the bulk correlation
length, a correlation function of the surface charge density can be
defined. It is checked, on the solvable models, that this correlation
function does have the generic long-range behaviour, decaying as the
inverse square distance,  predicted by macroscopic electrostatics. In 
the case of a two-component plasma (Coulomb fluid made of two species of
particles of opposite charges), the density correlation function on the 
boundary circle itself is conjectured to have a temperature-independent
behaviour, decaying as the -4 power of the distance.
\end{abstract}

\medskip

\noindent LPT Orsay 02-20

\newpage

\renewcommand{\theequation}{1.\arabic{equation}}
\setcounter{equation}{0}

\section{\bf Introduction}

This paper is dedicated to Jean-Pierre Hansen on the occasion of his
60th birthday. Although Jean-Pierre and I have coauthored only one 
paper \cite{HJS}, some 30 years ago, we share a longstanding interest
for Coulomb  fluids. Here is a contribution to this thriving domain of
research. It consists of a far from exhaustive minireview of previous
work on exactly solvable two-dimensional models of Coulomb fluids,
followed by an original part about a finite Coulomb fluid in a disc.

\subsection{\bf A brief review}
The classical (i.e. non-quantum) statistical mechanics of some 
two-dimensional models of 
Coulomb fluids is exactly solvable. These models have an intrisic 
interest: this is the only case of solvable models for a continuous
(i.e. not on a lattice) fluid, in more than one dimension. Furthermore, 
these models can be used as a testbench for a variety of generic 
properties of Coulomb fluids. For mimicking three-dimensional Coulomb
fluids in two dimensions, one must use the two-dimensional Coulomb
interaction which is logarithmic: the interaction energy between two
unit charges at a distance $r$ of each other is chosen as $\ln(L/r)$,
where $L$ is an arbitrary length which only fixes the zero of energy.
Indeed, this interaction is the two-dimensional solution of the Poisson
equation 
\begin{equation} \label{1.1}
\Delta\ln\frac{L}{r}=-2\pi\delta(\mathbf{r}).
\end{equation}
Of course, these logarithmic models do not describe ``real'' charged
particles, such as electrons, confined in a plane, which nevertheless
interact through the three-dimensional Coulomb potential $1/r$.

Two models are of special interest: The one-component plasma (OCP), or
jellium, is made of one species of particles of charge $q$, embedded in
a uniformly charged background of the opposite sign. The two-component
plasma (TCP), or Coulomb gas, is made of two species of particles, with
opposite charges $\pm q$. At the inverse temperature $\beta$, the
dimensionless coupling constant can be chosen as $\Gamma=\beta q^2$. It
has the remarkable property of being independent of the density, and
this results into a simple equation of state \cite{SP,HH} for both
models
\begin{equation} \label{1.2}
\beta p=(1-\frac{\Gamma}{4})n
\end{equation}
where $p$ is the pressure and $n$ is the density. 

In the case of the TCP, the model (with pure Coulomb interactions) 
becomes unstable againt the collapse of positive-negative pairs for
$\Gamma\geq 2$, and the equation of state (\ref{1.2}) is valid only for
$\Gamma\leq 2$. If some short-range repulsion (for instance hard cores
of diameter $a$) is introduced, the collapse at $\Gamma=2$ is 
suppressed and the temperature can be further lowered. For small values
of the dimensionless density $na^2$, the famous Kosterlitz-Thouless 
phase transition \cite{KT,AC,LLF} of infinite order occurs, between a 
conducting high-temperature phase and a dielectric low-temperature
phase, at a density-dependent temperature with a corresponding $\Gamma$
close to 4. 

Finite-size two-dimensional Coulomb fluids exhibit universal properties
\cite{SJ3} related to conformal invariance. This will not be reviewed
here.
 
For the special value $\Gamma=2$, the OCP \cite{AJ,J} and TCP \cite{CJ}
are fully exactly solvable: The thermodynamic functions and the
correlation functions are obtainable. 

For the OCP, it is convenient to define the Ursell function $U$ as
\begin{equation} \label{1.3}
U({\mathbf r},{\mathbf r}')=n^{(2)}({\mathbf r},{\mathbf r}')-n^2
\end{equation}
where $n$ is the one-body density and $n^{(2)}({\mathbf r},
{\mathbf r}')$ is the two-body density. At $\Gamma=2$, in the canonical
ensemble, the OCP maps on a system of free fermions in a magnetic field
and one finds 
\begin{equation} \label{1.4}
U({\mathbf r},{\mathbf r}')=-n^2\exp(-\pi n|{\mathbf r}-
{\mathbf r}'|^2).
\end{equation}
This correlation function has a fast Gaussian decay on a lengthscale of
the order of $(\pi n)^{-1/2}$. Several generic sum rules can be checked.
The zeroth and second moments of $U$ obey the two Stillinger-Lovett sum
rules \cite{SL} which express perfect screening of a charged particle 
of the fluid and perfect screening of an added infinitesimal test
charge, respectively. The fourth moment obeys the general expression
derived, in terms of the compressibility, by Vieillefosse and 
Hansen \cite{VH},
\begin{equation} \label{1.5}
n\left(\frac{\pi\Gamma}{4}\right)^2\int U({\mathbf r},{\mathbf r}')
|{\mathbf r}-{\mathbf r}'|^4{\mathrm d}{\mathbf r}=-\beta\left(
\frac{\partial p}{\partial n}\right)_T
\end{equation}
and the sixth moment obeys a recently derived expression \cite{KMST}
\begin{equation} \label{1.6}  
n^2\left(\frac{\pi\Gamma}{2}\right)^3\int U({\mathbf r},{\mathbf r}')
|{\mathbf r}-{\mathbf r}'|^6{\mathrm d}{\mathbf r}=
\frac{3}{4}(\Gamma-6)(8-3\Gamma).
\end{equation}

For the TCP, the Ursell functions $U$ depend on the species of the two
particles which are involved. They must be defined as
\begin{equation} \label{1.7}
U_{ss'}({\mathbf r},{\mathbf r}')=n_{ss'}^{(2)}({\mathbf r},
{\mathbf r}')-n_sn_{s'} 
\end{equation}
where $s,s'=\pm 1$ denote the signs of the particles, $n_+=n_-=n/2$
are the one-body densities of each species, $n$ the total one-body
density, and $n_{ss'}^{(2)}({\mathbf r},{\mathbf r}')$ is the two-body
density for particles of species $s$ and $s'$, respectively.  At 
$\Gamma=2$, in the grand canonical ensemble, the TCP maps on a system of
free fermions. A control parameter is a properly rescaled fugacity $m$,
which has the dimension of an inverse length. Although the density
diverges at $\Gamma=2$, the Ursell functions remain finite. They are
\begin{subequations} \label{1.8}
\begin{equation} \label{1.8a}
U_{ss}({\mathbf r},{\mathbf r}')=-\left(\frac{m^2}{2\pi}\right)^2
[K_0(m|{\mathbf r}-{\mathbf r}'|)]^2 
\end{equation}
\begin{equation} \label{1.8b}
U_{s\,-s}({\mathbf r},{\mathbf r}')=\left(\frac{m^2}{2\pi}\right)^2
[K_1(m|{\mathbf r}-{\mathbf r}'|)]^2
\end{equation}
\end{subequations}
where $K_0$ and $K_1$ are modified Bessel functions, which have an
exponential decay; the correlation length is $1/(2m)$. The short-
distance behaviours of these Ursell functions confirm general 
predictions of Hansen and Viot \cite{HV}. The pair distribution 
function for particles of opposite signs, dominated at short distances 
by the Bolzmann factor of the Coulomb potential, should behave like 
$|{\mathbf r}-{\mathbf r}'|^{-\Gamma}$; the 
$|{\mathbf r}-{\mathbf r}'|^{-2}$ behaviour of $U_{s\,-s}$ fits with
this form. However, although the pair distribution function for
particles of the same sign should behave like 
$|{\mathbf r}-{\mathbf r}'|^{\Gamma}$  
for $\Gamma <1$, this repulsive behaviour has been predicted to be 
weakened into $|{\mathbf r}-{\mathbf r}'|^{2-\Gamma}$ for $1<\Gamma<2$.
because of screening by a third particle; the logarithmic behaviour of 
the Bessel function $K_0$ in $U_{ss}$ can be considered as a limiting 
case when $\Gamma=2$. 

The Ursell functions for the TCP can be combined into a charge
correlation function
\begin{equation} \label{1.9}
U_{\rho}=2q^2(U_{++}-U_{-+})
\end{equation}
and a density correlation function
\begin{equation} \label{1.10}
U_n=2(U_{++}+U_{-+}).
\end{equation}
Several generic sum rules can be checked on these correlation
functions. The zeroth and second moment of the charge correlation
function (\ref{1.9}) obey the Stillinger-Lovett sum rules. The zeroth
moment of the density correlation function (\ref{1.10}) obeys the usual
compressibility sum rule, while its second moment obeys a recently
dicovered sum rule \cite{J2,JKS} which seems to be specific to the
two-dimensional point-particle TCP:
\begin{equation} \label{1.11}
\int U_n({\mathbf r},{\mathbf r}')|{\mathbf r}-{\mathbf r}'|^2
{\mathrm d}{\mathbf r}=\frac{1}{12\pi[1-(\Gamma/4)^2]}.
\end{equation}

Up to two years ago,there were exact results only at $\Gamma=2$. Then a
major breakthrough occurred: \v{S}amaj et al. \cite{ST} succeeded in 
deriving the thermodynamic properties of the TCP in the whole range of
stability of the model, $\Gamma<2$. This was achieved by using a mapping
on the sine-Gordon field theory and results known for the
latter. The surface tension along a wall could also be obtained in the
cases of an ideal conductor wall \cite{SJ} and of an ideal dielectric 
wall \cite{S}. However, although results have been obtained for the
asymptotic behaviour of the correlation functions \cite{SJ2,SJ3}, there
are no simple expressions for these correlation functions. In the
following, we shall consider only the simple case $\Gamma=2$, and the
weak-coupling (high-temperature) limit $\Gamma\rightarrow 0$.

At $\Gamma=2$, many exact results are available for the OCP and the TCP 
in a variety of geometries with walls or on a curved surface.
These results will not be described in detail, the interested reader 
might look on cond-mat for Forrester, Jancovici, T\'ellez \ldots.
Here, we only review the case of a Coulomb fluid living in a
half-plane bounded by a rectilinear hard wall. This hard wall can be
taken as the $y$ axis, and the fluid is supposed to occupy the
half-plane $x\geq 0$. The position ${\mathbf r}$ of a particle is
defined by its Cartesian coordinates $x$ and $y$. Now, in the
definitions of the Ursell functions, it must be understood that the 
one-body densities are functions of the distance to the wall, i.e. 
in (\ref{1.3}) $n^2$ must be replaced by $n(x)n(x')$, and in (\ref{1.7}) 
$n_sn_{s'}$ must be replaced by $n_s(x)n_{s'}(x')$. The Ursell functions 
depend on $x$, $x'$, and $|y-y'|$. 

For this half-plane geometry, a generic behaviour of the charge
correlations near the wall \cite{J3,J4,J5} results from the assumption 
that the fluid is a conductor obeying the laws of macroscopic
electrostatics. The charge correlation function along the wall now is
long-ranged, with only an algebraic asymptotic decay 
\begin{equation} \label{1.12}
U_{\rho}(x,x',|y-y'|)\sim \frac{f(x,x')}{(y-y')^2}
\end{equation}
when $|y-y'|$ is large compared to the microscopic scale (the bulk
correlation length). $f(x,x')$ is a function which is localized near the
wall (it has a fast decay as $x$ or $x'$ increase beyond the microscopic 
scale), and $f$ obeys the sum rule
\begin{equation} \label{1.13}
\beta\int_0^{\infty}{\mathrm d}x \int_0^{\infty}{\mathrm d}x'f(x,x')
=-\frac{1}{2\pi^2}.
\end{equation} 
(\ref{1.12}) and (\ref{1.13}) can be reexpressed by writing that there
is a surface charge density $\sigma(y)$ with a correlation function
obeying
\begin{equation} \label{1.13bis}
\beta<\sigma(y)\sigma(y')>=-\frac{1}{2\pi^2(y-y')^2}
\end{equation}
At $\Gamma=2$ and in the limit $\Gamma\rightarrow 0$, along a hard wall,
the charge correlation functions $q^2U$ of the OCP \cite{J3} and
$U_{\rho}$ of the TCP \cite{SJ3} do have an asymptotic behaviour
in agreement with (\ref{1.13bis}).  

For this half-plane geometry, it has been observed \cite{SJ3} that 
the density correlation function of the TCP near the wall is also
long-ranged and that $U_n(x=0,x'=0,|y-y'|)$ has the same asymptotic 
behaviour at $\Gamma=2$ and as $\Gamma\rightarrow 0$: 
\begin{equation} \label{1.14}
U_n(x=0,x'=0,|y-y'|)\sim\frac{1}{2\pi^2(y-y')^4}. 
\end{equation}
It is tempting to conjecture that (\ref{1.14}) is valid at any
temperature.

\subsection{\bf The disc geometry}
We now come to the original part of the present paper. In the above, it
was always found that the laws of macroscopic electrostatics about
charge correlations were satisfied by the exactly solvable
two-dimensional models, when the lengths under consideration are large
compared to the microscopic scale. Recently, however, a counterexample
was found \cite{J6}. A short-circuited circular capacitor was
considered: A two-dimensional Coulomb fluid fills a disk of radius $R_1$
and the infinite region outside a concentric circle of larger radius
$R_2$, while the annulus between $R_1$ and $R_2$ is empty, and the two
filled regions are allowed to freely exchange charged particles. This
geometry is exactly solvable at $\Gamma=2$ for both the OCP and the
TCP. The charge $Q$ on the inner disk fluctuates. Even for macroscopic
values of $R_1$, $R_2$, with $R_2/R_1>1$, it was found that the variance
$<Q^2>-<Q>^2$ \emph{differs} from the value predicted by using linear
response theory and macroscopic electrostatics    
\begin{equation} \label{1.15}
\beta(<Q^2>-<Q>^2)=\frac{1}{\ln\frac{R_2}{R_1}}.
\end{equation}
This, at first sight surprising, disagreement can be explained, on
second thought, when it is noted that the fluctuations (\ref{1.15})
involve only a small number of particles, because, at $\Gamma=2$, 
$(<Q^2>-<Q>^2)/q^2$ is of order unity. Although, for one disc alone, 
i.e. in the limit $R_2\rightarrow\infty$, no charge fluctuations are
found for the solvable models (except in one very special case), in
agreement with the limit $R_2\rightarrow\infty$ in (\ref{1.15}), the
failure of (\ref{1.15}) for $R_2/R_1$ finite casts a reasonable doubt on
the predictions of macroscopic electrostatics in the disc geometry. The
subject of the present paper is to investigate the surface charge 
correlations of the OCP and TCP in a disc, at $\Gamma=2$. It will be
shown that macroscopic electrostatics does hold for the present problem.

Thus, we consider a Coulomb fluid in a disc of radius $R$, bounded by a
hard wall. It is convenient to put the origin at the centre of the disc,
and to use polar coordinates $(r,\varphi)$. The disc may be either
insulated, with for simplicity a vanishing total charge, or grounded. In
both cases, from linear response theory and macroscopic electrostatics
\cite{J5} one finds that the correlation function of the surface charge 
density $\sigma(\varphi)$ is given by
\begin{equation} \label{1.16}
\beta<\sigma(\varphi)\sigma(\varphi')>=-\frac{1}
{2\pi^2[2R\sin(\theta/2)]^2}
\end{equation}
where $\theta=\varphi-\varphi'$ is the angular distance between the two
points.  Actually, the correlation function in (\ref{1.16}) is
defined from the microscopic charge Ursell function 
$U_{\rho}(r,r',|\theta|)$, which, for $R$ and $R|\theta|$ large compared
to the microscopic scale, is localized near the boundary circle 
(it has a fast decay as $x=R-r$ or $x'=R-r'$ increase beyond the 
microscopic scale). The relation is
\begin{equation} \label{1.17}
<\sigma(\varphi)\sigma(\varphi')>=\int_0^{\infty}{\mathrm d}x
\int_0^{\infty}{\mathrm d}x'U_{\rho}(r,r',|\theta|).
\end{equation} 
Also, possible oscillations as a function of $\theta$ on a microscopic
scale are washed out in the definition of the surface charge density.
(\ref{1.16}) is a generalization of the rectilinear wall case 
(\ref{1.13bis}), which is retrieved from (\ref{1.16}) in
the limit $R\rightarrow\infty$, $\theta\rightarrow 0$, at a fixed value
of $R\theta$ which becomes $y-y'$.

In section 2, it will be shown that (\ref{1.16}) is obeyed in the case
of the OCP at $\Gamma=2$. In section 3, it will be shown that
(\ref{1.16}) is obeyed in the case of the TCP at $\Gamma=2$. A
generalization to the disc geometry of the density correlation  
function on the wall for the TCP (\ref{1.14}) will also be presented.
The high-temperature limit $\Gamma\rightarrow 0$ will be briefly
reviewed in section 4.

\renewcommand{\theequation}{2.\arabic{equation}}
\setcounter{equation}{0}

\section{{\bf One-component plasma at} ${\mathbf \Gamma=2}$ {\bf in a
disc}}

The canonical ensemble is used. There are $N$ particles of charge $q$ in
a disc of radius $R$. Thus, the average density is $n=N/(\pi R^2)$. A
uniformly charged background insures that the total charge vanishes. It
is convenient to choose the unit of length as $R/\sqrt{N}$. In these
units, $\pi n=1$. The general formalism \cite{J3} for the OCP in a disk
at $\Gamma=2$ expresses the density Ursell function 
$U(r,r',|\theta|)=n^{(2)}(r,r',|\theta|)-n(r)n(r')$ in terms of an
auxiliary function 
\begin{equation} \label{2.1}
K(w)=\sum_{l=0}^{N-1}\frac{w^l}{\gamma(l+1,N)} 
\end{equation}
where $w=rr'\exp({\mathrm i}\theta)$ and $\gamma(l+1,N)$ is an 
incomplete gamma function \cite{E}, as
\begin{equation} \label{2.2}
U(r,r',|\theta|)=-n^2\exp(-r^2-r'^2)|K(w)|^2.
\end{equation}
We are interested in the behaviour of (\ref{2.2}) when $r$ and $r'$ are 
 close to $R$ and $R\theta$ is much larger than the microscopic scale 
$n^{-1/2}$, i.e. for $N$ large and a non-zero value of $\theta$.

A related problem has been previously solved by Choquard et
al. \cite{Ch}. They considered the simpler case when the confining 
circular background extends well beyond the circular blob formed by the
particles. Then, in (\ref{2.1}), the incomplete gamma function is
replaced by the complete one $\Gamma(l+1)=l!$ and the sum itself can be
expressed in terms of another incomplete gamma function, from which the
surface behaviour of $U$ could be obtained (the surface behaviour of $U$
has also been obtained \cite{FJ} in the case of an elliptical blob of
particles). In our present case, since for $r$ and $r'$ close to $R$ the
sum in (\ref{2.1}) can be seen to be dominated by values of $l$ close to
$N$, we use the asymptotic expression \cite{E} $\gamma(l+1,l)\sim
(1/2)l!$. Thus our $K$ is just twice the one of Choquard et al. Although
we have not been able to make a rigorous mathematical proof, the
validity of this procedure has been checked on Mathematica, in the cases
$r=r'=R$, $\theta=\pi/2$ or $\theta=\pi$, by numerical evaluations of a
sum equivalent to (\ref{2.1}).

Following Choquard et al., we can now express our $K$ as \cite{E}
\begin{equation} \label{2.3}
K(w)=2\sum_{l=0}^{N-1}\frac{w^l}{l!}=
\frac{2}{(N-1)!}\exp(w)\Gamma(N,w) 
\end{equation}
where $\Gamma(N,w)$ is the incomplete gamma function
\begin{equation} \label{2.4}
\Gamma(N,w)=\int_w^{\infty}{\mathrm d}t{\mathrm e}^{-t}t^{N-1}.
\end{equation}
The asymptotic form of (\ref{2.4}), in the present case of 
$|w-N|\gg\sqrt N$ and $N\rightarrow\infty$, can be obtained by 
rewriting (\ref{2.4}) as
\begin{equation} \label{2.5} 
\Gamma(N,w)=\int_w^{\infty}{\mathrm d}t\frac{t}{N-1-t}\frac{{\mathrm d}}
{{\mathrm d}t}(t^{N-1}{\mathrm e}^{-t})
\end{equation} 
and integrating by parts, with the result 
\begin{equation} \label{2.6}
\frac{\exp(w)}{(N-1)!}\Gamma(N,w)=\frac{w^N}{(N-1)!(w-N+1)}
[1+O(\frac{1}{N})]
\end{equation}
in agreement with an asymptotic expansion by Tricomi (see \cite{E}). In
the case of two particles on the circle of radius $R$, since in our
units $R^2=N$, $w=N\exp({\mathrm i}\theta)$, and from (\ref{2.3}) 
and (\ref{2.6}) one finds
\begin{equation} \label{2.7}
{\mathrm e}^{-N}K(N{\mathrm e}^{{\mathrm i}\theta})\sim
\frac{2{\mathrm e}^{-N}N^N}{N!}\frac{{\mathrm e}^{{\mathrm i}N\theta}}
{{\mathrm e}^{{\mathrm i}\theta}-1}.
\end{equation}
Using Stirling's formula for $N!$ in (\ref{2.7}) gives
\begin{equation} \label{2.8}
{\mathrm e}^{-N}K(N{\mathrm e}^{{\mathrm i}\theta})\sim
\left(\frac{2}{\pi N}\right)^{1/2}\frac
{{\mathrm e}^{{\mathrm i}N\theta}}{{\mathrm e}^{{\mathrm i}\theta}-1}.
\end{equation}
Using (\ref{2.8}) in (\ref{2.2}) and reestablishing an arbitrary unit of
length such that $\pi n$ is no longer 1, gives the correlation function 
on the circle 
\begin{equation} \label{2.9}
U(R,R,\theta)\sim-n\frac{2}{\pi^2}\frac{1}
{\left(2R\sin\frac{\theta}{2}\right)^2}.
\end{equation}
In the case when the two particles are at small distances $x$ and $x'$
from the boundary,  minor modifications after (\ref{2.6}) give, in the 
large-$N$ limit
\begin{equation} \label{2.10}
U(r,r',\theta)\sim -n\frac{2}{\pi^2}\frac{\exp(-2\pi nx^2-2\pi nx'^2)}
{\left(2R\sin\frac{\theta}{2}\right)^2}.
\end {equation}

Integrating $U_{\rho}=q^2U$ with respect to $x$ and $x'$ does give a
correlation function of the surface charge density of the form
(\ref{1.16}) (here $\beta q^2=2$), in agreement with macroscopic
electrostatics.

The case of a rectilinear hard wall \cite{J3} is retrieved from 
(\ref{2.10}) by taking the limit $R\rightarrow\infty$, 
$\theta\rightarrow 0$, at a fixed value of $R\theta$ which becomes
$y-y'$. Then, in agreement with (\ref{1.12}),
\begin{equation} \label{2.11} 
U_{\rho}(x,x',|y-y'|)\sim
-q^2n\frac{\exp(-2\pi nx^2-2\pi nx'^2)}{(y-y')^2}
\end{equation}
and (\ref{1.13bis}) is satisfied.

In the derivation of the present results, we have replaced 
$\gamma(l+1,N)$ by $(1/2)l!$ without a rigorous mathematical
justification. However, the validity of (\ref{2.9}) and (\ref{2.10}) is
supported by several checks. First, the correct result (\ref{2.11}) 
was retrieved in the limiting case of a rectilinear wall. Second, using
the exact expression (\ref{2.1}), Choquard et al. \cite{Ch2} have
shown without approximations that, in the large-$N$ limit,
\begin{equation} \label{2.13}
-\frac{1}{2\pi R^2}\int{\mathrm d}{\mathbf r}{\mathrm d}{\mathbf r}'
|{\mathbf r}-{\mathbf r}'|^2U({\mathbf r},{\mathbf r}')=\frac{1}{\pi}
\end{equation}
and that the surface contribution to (\ref{2.13}) is $1/(2\pi)$; this is
in agreement with what is obtained by using for $U$ near the surface 
the expression (\ref{2.10}).
   
\renewcommand{\theequation}{3.\arabic{equation}}
\setcounter{equation}{0}

\section{{\bf Two-component plasma at} ${\mathbf \Gamma=2}$ {\bf in a
disc}}

The particles are confined by a hard wall in a disc of radius $R$. A
grand canonical ensemble restricted to neutral configurations is used. 
Both species of particles, of respective charges $q$ and $-q$, have the
same rescaled fugacity $m$. The general formalism \cite{CJ} expresses
the Ursell functions $U_{ss'}(r,r',|\theta|)$ in terms of Green
functions $G_{ss'}({\mathbf r},{\mathbf r}')$ as
\begin{equation} \label{3.1}
U_{ss'}(r,r',|\theta|)=-ss'm^2|G_{ss'}({\mathbf r},{\mathbf r}')|^2.
\end{equation}
Because of the symmetry between positive and negative particles, we only
need $G_{++}$ and $G_{-+}$ which, for $r,r'<R$, are determined by
\begin{equation} \label{3.2}
(m^2-\Delta)G_{++}({\mathbf r},{\mathbf r}')=m\delta
({\mathbf r}-{\mathbf r}')
\end{equation}
and
\begin{equation} \label{3.3}
G_{-+}({\mathbf r},{\mathbf r}')=-\frac{\exp({\mathrm i}\varphi)}{m}
\left(\frac{\partial}{\partial r}+\frac{{\mathrm i}}{r}\frac{\partial}
{\partial\varphi}\right)G_{++}({\mathbf r},{\mathbf r}').
\end{equation}
In infinite space the solution of (\ref{3.2}) is $[m/(2\pi)]K_0
(m|{\mathbf r}-{\mathbf r}'|)$. In a disc, it is appropriate to use
polar coordinates and to write the solution as an expansion of the form
\begin{equation} \label{3.4}
G_{++}(r,\varphi;r',\varphi')=\frac{m}{2\pi}\sum_{l=-\infty}^{\infty}
[I_l(mr')K_l(mr)+a_lI_l(mr')I_l(mr)]\exp[{\mathrm i}l(\varphi-\varphi')]
\ \ (r'<r<R)
\end{equation}
where the first term in the sum is the expansion of $[m/(2\pi)]K_0
(m|{\mathbf r}-{\mathbf r}'|)$ and the second term is a ``reflected''
contribution due to the wall; $I_l$ and $K_l$ are modified Bessel
functions and $a_l$ is a coefficient to be determined by the boundary
conditions. (\ref{3.3}) gives
\begin{eqnarray} \label{3.5}
&&G_{-+}(r,\varphi;r',\varphi')=\frac{m}{2\pi}\sum_{l=-\infty}^{\infty}
[I_l(mr')K_{l+1}(mr) \nonumber \\
&&-a_lI_l(mr')I_{l+1}(mr)]
\exp[{\mathrm i}(l+1)\varphi-{\mathrm i}l\varphi'] \ \ (r'<r<R)
\end{eqnarray}
For $r'<R$, $r>R$, $m$ must be replaced by 0 in (\ref{3.2}), and, as
functions of ${\mathbf r}$, $G_{++}$ depends only on 
$z=r\exp({\mathrm i}\varphi)$, $G_{-+}$ depends only on $\bar{z}=
r\exp(-{\mathrm i}\varphi)$, and they must vanish at infinity. At $r=R$, 
$G_{++}$ and $G_{-+}$ must be continuous. These conditions impose that
the $l\geq 0$ terms in $G_{++}(R,\varphi;r',\varphi')$ and the terms
$l<0$  in $G_{-+}(R,\varphi;r',\varphi')$ vanish. Therefore, the
coefficient $a_l$ in (\ref{3.4}) and (\ref{3.5}) is
\begin{equation} \label{3.6} 
a_l=-\frac{K_l(mR)}{I_l(mR)}\ {\mathrm if}\ l\geq 0,\ \ a_l=
\frac{K_{l+1}(mR)}{I_{l+1}(mR)}\ {\mathrm if}\ \l<0.
\end{equation}
We are interested in the behaviours of (\ref{3.4}) and (\ref{3.5}) when
the disc is much larger that the bulk correlation length ($mR\gg 1$), 
$\theta=\varphi-\varphi'$ has a fixed (non-zero) value, and $r$ and $r'$
are close to $R$. 
 
Let us start with the case when the two points are on the
disc. In this limiting case, when the Wronskian relation \cite{E}  
$I_l(mR)K_{l+1}(mR)+I_{l+1}(mR)K_l(mR)=1/(mR)$ is taken into
account, and after $l$ has been changed into $-l$, (\ref{3.4}) and
(\ref{3.6}) give 
\begin{equation} \label{3.7}
G_{++}(R,\varphi;R,\varphi')=\frac{1}{2\pi R}\sum_{l=1}^{\infty}
\frac{I_l(mR)}{I_{l-1}(mR)}\exp[{-\mathrm i}l(\varphi-\varphi')]
\end{equation}
while, when that same Wronskian relation is used, (\ref{3.5}) and
(\ref{3.6}) give
\begin{equation} \label{3.8}
G_{-+}(R,\varphi;R,\varphi')=\frac{1}{2\pi R}\sum_{l=0}^{\infty}
\exp[{\mathrm i}(1+l)\varphi-{\mathrm i}l\varphi')].
\end{equation}
In the large-$mR$ limit, using in (\ref{3.7}) the large-argument
asymptotic expansions of the Bessel functions \cite{AS} give, up to 
order $1/R^2$,
\begin{equation} \label{3.9}
\frac{I_l(mR)}{I_{l-1}(mR)}=1+\frac{1-2l}{2mR}+\frac{3-8l+4l^2}{8(mR)^2}
+\ldots\ .
\end{equation}
Although the sums on $l$ which appear in (\ref{3.8}), and in (\ref{3.7})
when (\ref{3.9}) is used, are not convergent, they can be given
a meaning in the sense of distributions; the obvious recipe is to insert
a convergence factor $p^l$, with $|p|<1$, in each term, and to take the
limit $p\rightarrow 1$ after the summation. One finds, up to order
$1/R^3$,
\begin{eqnarray} \label{3.10}  
G_{++}(R,\varphi;R,\varphi')&=&\frac{1}{2\pi R}\,
\frac{{\mathrm e}^{-{\mathrm i}\theta}}
{1-{\mathrm e}^{-{\mathrm i}\theta}}
\left[1-\frac{1}{2mR}\,\frac{1+{\mathrm e}^{-{\mathrm i}\theta}}
{1-{\mathrm e}^{-{\mathrm i}\theta}} \right. \nonumber \\ 
&+&\left.\frac{1}{8(mR)^2}\,\frac{-1+6{\mathrm e}^{-{\mathrm i}\theta}
+3{\mathrm e}^{-2{\mathrm i}\theta}}{(1-{\mathrm e}^
{-{\mathrm i}\theta})^2}+\ldots\right] 
\end{eqnarray}
where $\theta=\varphi-\varphi'$, and
\begin{equation} \label{3.11}
G_{-+}(R,\varphi;R,\varphi')=\frac{{\mathrm e}^{{\mathrm i}\varphi}}
{2\pi R}\,\frac{1}{1-{\mathrm e}^{{\mathrm i}\theta}}.
\end{equation}
It might be noted that the expression (\ref{3.11}) is exact, without any
large-$mR$ expansion. Finally, using (\ref{3.10}) or (\ref{3.11}) in
(\ref{3.1}) gives
\begin{equation} \label{3.12}
U_{++}(R,R,|\theta|)=-\frac{m^2}
{4\pi^2\left(2R\sin\frac{\theta}{2}\right)^2}
+\frac{1}{4\pi^2\left(2R\sin\frac{\theta}{2}\right)^4}+\ldots
\end{equation}
up to order $1/R^4$, and
\begin{equation} \label{3.13}
U_{-+}(R,R,|\theta|)=\frac{m^2}
{4\pi^2\left(2R\sin\frac{\theta}{2}\right)^2}.
\end{equation}

More generally, if the two points are at small distances $x$ and $x'$
from the boundary, using the large-argument asymptotic expansions of the
Bessel functions in (\ref{3.4}) and (\ref{3.5}) (where the first term in
the sum, which comes from the expansion of $K_0$, gives a short-ranged
contribution which can be omitted) gives, when only the leading term is
kept,
\begin{equation} \label{3.14}
G_{++}(R-x,\varphi;R-x',\varphi')\sim\frac{{\mathrm e}^{-m(x+x')}}
{2\pi R}\,\frac{{\mathrm e}^{-{\mathrm i}\theta}}
{1-{\mathrm e}^{-{\mathrm i}\theta}}
\end{equation}
and
\begin{equation} \label{3.15}
G_{-+}(R-x,\varphi;R-x',\varphi')\sim\frac{{\mathrm e}^{-m(x+x')+
{\mathrm i}\varphi}}{2\pi R}\,\frac{1}
{1-{\mathrm e}^{{\mathrm i}\theta}}.
\end{equation}
The Ursell functions become
\begin{equation} \label{3.16}
U_{++}(R-x,R-x',|\theta|)\sim-\frac{m^2{\mathrm e}^{-2m(x+x')}}
{4\pi^2\left(2R\sin\frac{\theta}{2}\right)^2}
\end{equation}
and
\begin{equation} \label{3.17}
U_{-+}(R-x,R-x',|\theta|)\sim\frac{m^2{\mathrm e}^{-2m(x+x')}}
{4\pi^2\left(2R\sin\frac{\theta}{2}\right)^2}.
\end{equation}

The charge correlation function is
\begin{equation} \label{3.18}
U_{\rho}(R-x,R-x',|\theta|)=2q^2[U_{++}(R-x,R-x',|\theta|)-
U_{-+}(R-x,R-x',|\theta|)].
\end{equation}
Using (\ref{3.16}) and (\ref{3.17}) in (\ref{3.18}) and integrating it
on $x$ and $x'$ does give a correlation function of the surface charge
density of the form (\ref{1.16}) (here $\beta q^2=2$), in agreement with 
macroscopic electrostatics.
 
The density correlation function on the wall is
\begin{equation} \label{3.19}
U_n(R,R,|\theta|)=2[U_{++}(R,R,|\theta|)+U_{-+}(R,R,|\theta|)].
\end{equation}
In (\ref{3.19}), the leading contribution to $U_{++}$ is cancelled by
$U_{-+}$ and it is necessary to use the full expansion (\ref{3.12}),
together with (\ref{3.13}) for obtaining the leading term of $U_n$ as
\begin{equation} \label{3.20}  
U_n(R,R,|\theta|)\sim\frac{1}
{2\pi^2\left(2R\sin\frac{\theta}{2}\right)^4}.
\end{equation}
(\ref{3.20}) is the generalization to the case of a circular wall of the
rectilinear wall result (\ref{1.14}).

A mathematical justification of some of the heuristic steps used in the 
above derivations is given in the Appendix.

\renewcommand{\theequation}{4.\arabic{equation}}
\setcounter{equation}{0}

\section{{\bf High-temperature limit} ${\mathbf \Gamma\rightarrow 0}$
{\bf in a disc}}

For both the OCP and the TCP, the high-temperature limit
$\Gamma\rightarrow 0$ is described by the Debye-H\"uckel theory.
The case of a disc has already been investigated by Choquard 
et al. \cite{Ch3}. They have shown that, when $x=R-r$ and $x'= R-r'$ are
small, the charge correlation function is\footnote{Actually, Choquard 
et al. have given, instead of (\ref{4.1}), a more general expression
involving the arbitrary length $L$ in the logarithmic potential
$\ln(L/r)$. We have argued \cite{J6} that, in two dimensions, there
should be no fluctuations of the total charge on the disc, even in a
grand canonical ensemble, and that this condition imposes that the limit
$L\rightarrow\infty$ should be taken. Then, the expression of Choquard
et al. reduces to (\ref{4.1}).}
\begin{equation} \label{4.1}
U_{\rho}(r,r',|\theta|)\sim-\frac{q^2n{\mathrm e}^{-\kappa (x+x')}}
{\pi\left(2R\sin\frac{\theta}{2}\right)^2}
\end{equation}
where $\kappa=(2\pi\beta q^2n)^{1/2}$ is the the inverse Debye 
length. Integrating (\ref{4.1}) on $x$ and $x'$ does give a
correlation function of the surface charge density of the form
(\ref{1.16}), in agreement with macroscopic electrostatics.

In the case of the TCP, the density correlation function $U_n$ is also
of interest. Since the Debye-H\"uckel theory gives a vanishing result
(the contributions from the Debye-H\"uckel $U_{ss}$ and $U_{s\,-s}$
cancel each other), it is necessary to go beyond the Debye-H\"uckel
theory and to take into account the next term in the renormalized Mayer
expansion \cite{JKS,SJ2}, which is proportional to the square of the
Debye-H\"uckel $U_{\rho}$. One finds
\begin{equation} \label{4.2}   
U_n(R-x,R-x',|\theta|)\sim\frac{{\mathrm e}^{-2\kappa (x+x')}}
{2\pi^2\left(2R\sin\frac{\theta}{2}\right)^4}.
\end{equation}
On the wall itself, (\ref{4.2}) is the same as in the case (\ref{3.20})
of $\Gamma=2$.

\renewcommand{\theequation}{5.\arabic{equation}}
\setcounter{equation}{0}

\section{\bf Conclusion}

Since, in two dimensions, we could have some doubts about the validity
of macroscopic electrostatics, which is used for deriving the
correlation function (\ref{1.16}) of the surface charge density for a
Coulomb fluid in a disc, we have checked this correlation function on
exactly solvable models, the OCP and TCP at $\Gamma=2$. That correlation 
function had also been obtained in the Debye-H\"uckel 
theory \cite{Ch3}, i.e. in the weak-coupling $\Gamma\rightarrow 0$ 
limit. Thus, these calculations indicate that macroscopic electrostatics
is valid for describing the correlation function of the surface charge
density. Incidentally, this correlation function (times $\beta$) can be
formally written as a Fourier series
\begin{equation} \label{5.1}
\beta<\sigma(\varphi)\sigma(\varphi')>=-\frac{1}{2\pi^2[2R\sin(\theta/2)]^2}
=\frac{1}{2\pi^2R^2}\sum_{l=1}^{\infty}l\cos(l\theta). 
\end{equation}
This Fourier series has no $l=0$ term, another indication that the total
charge has no fluctuations.

In the case of a TCP, the density correlation function on the wall has
the same form (\ref{3.20}) at $\Gamma=2$ and in the high-temperature
limit $\Gamma\rightarrow 0$. It is tempting to conjecture that this form
is valid at any temperature, at least in the range $\Gamma\leq 2$. A
similar universality of (\ref{1.14}) had been conjectured \cite{SJ3} in
the case of a rectilinear wall. Proving (or disproving) these
conjectures is an open problem.

\renewcommand{\theequation}{A.\arabic{equation}}
\setcounter{equation}{0}

\section*{\bf Appendix}

In section 3, large-\emph{argument} expansions of Bessel functions have
been used, although these expansions have been inserted into infinite 
series involving also large values of the index $l$. Furthermore, these
series in $l$ have been made to converge only through the introduction
of an ad hoc convergence factor. In the present Appendix, some
justification is given. For simplicity, only one case is considered, the
calculation of $G_{-+}(R,\varphi;r'=R-x',\varphi')$. 

 From (\ref{3.5}), (\ref{3.6}), and the Wronskian relation,
\begin{equation} \label{A.1}
G_{-+}(R,\varphi;r',\varphi')=\frac{1}{2\pi R}\sum_{l=0}^{\infty}
\frac{I_l(mr')}{I_l(mR)}
\exp[{\mathrm i}(1+l)\varphi-{\mathrm i}l\varphi')].
\end{equation}
Using the \emph{uniform} asymptotic expansions of the Bessel functions
\cite{AS}, appropriate here since the arguments $mr'$ and $mR$ are large 
but the index $l$ may also be large, gives after some algebra 
\begin{equation} \label{A.2}
G_{-+}(R,\varphi;r',\varphi')\sim\frac{{\mathrm e}^
{{\mathrm i}\varphi}}{2\pi R}\sum_{l=0}^{\infty}
\exp[-mx'(1+\frac{l^2}{m^2R^2})^{1/2}]\exp({\mathrm i}l\theta)
\end{equation}
where $\theta=\varphi-\varphi'$. We are interested in the 
$mR\rightarrow\infty$ limit of (\ref{A.2}), for a fixed value of $mx'$
and $\theta\neq 0$. For obtaining this limit, the Borel summation method
is used. The sum in (\ref{A.2}) can be written as 
\begin{equation} \label{A.3}
S=\int_0^{\infty}{\mathrm d}t\exp(-t)f(t)
\end{equation}
where
\begin{equation} \label{A.4}
f(t)=\sum_{l=0}^{\infty}\exp[-mx'(1+\frac{l^2}{m^2R^2})^{1/2}]
\exp({\mathrm i}l\theta)\frac{t^l}{l!}.
\end{equation}
Since the sum (\ref{A.4}) is absolutely convergent, the limit and the
sum can be interchanged, giving
\begin{equation} \label{A.5}.
\lim_{mR\rightarrow\infty}f(t)=\exp(-mx')\exp(t{\mathrm e}^
{{\mathrm i}\theta})
\end{equation}
Using the limit (\ref{A.5}) in (\ref{A.3}) gives the large-$mR$
behaviour
\begin{equation} \label{A.6}
G_{-+}(R,\varphi;R-x',\varphi')\sim\frac{{\mathrm e}^
{-mx'+{\mathrm i}\varphi}}{2\pi R}\,\frac{1}
{1-{\mathrm e}^{{\mathrm i}\theta}}
\end{equation}
in agreement with (\ref{3.15}).

\section*{\bf Acknowledgements}

I have benefited of stimulating discussions with F. van Wijland. The use
of the Borel summation method in the Appendix was suggested by K. Chadan.
I am indebted to L. \v{S}amaj for a critical reading of the manuscript.


\newpage

\begin{thebibliography}{99}

\bibitem{HJS} Hansen J P, Jancovici B, and Schiff D 1972 
{\it Phys. Rev. Lett.} {\bf 29} 991

\bibitem{SP} Salzberg A and Prager S 1963 {\it J. Chem. Phys.} {\bf 38}
2587

\bibitem{HH} Hauge E H and Hemmer P C 1971 {\it Phys. Norvegica} {\bf 5}
209 

\bibitem{KT} Kosterlitz J M and Thouless D J 1973 {\it J. Phys. C} 
{\bf 6} 1181

\bibitem{AC} Alastuey A and Cornu F 1992 {\it J. Stat. Phys.} {\bf 66}
165 and references quoted there

\bibitem{LLF} Levin Y, Xiao-jun L, and Fisher M E 1994 
{\it Phys. Rev. Lett.} {\bf 73} 2716 and references quoted there
 
\bibitem{SJ3} \v{S}amaj L and Jancovici B 2002 {\it J. Stat. Phys.} 
{\bf106} 323 and references quoted there

\bibitem{AJ} Alastuey A and Jancovici B 1981 {\it J. Phys.(France)} 
{\bf 42} 1

\bibitem{J} Jancovici B 1981 {\it Phys. Rev. Lett.} {\bf 46} 386

\bibitem{CJ} Cornu F and Jancovici B 1989 {\it J. Chem. Phys.} {\bf 90}
2444 

\bibitem{SL} Stillinger F H and Lovett R 1968 {\it J. Chem. Phys.} 
{\bf 49} 1991

\bibitem{VH} Vieillefosse P and Hansen J P 1975 {\it Phys. Rev. A}
{\bf 12} 1106

\bibitem{KMST} Kalinay P, Mark\v{o}s P, \v{S}amaj L, and 
Trav\v{e}nec I, 2000 {\it J. Stat. Phys.} {\bf 98} 639

\bibitem{HV} Hansen J P and Viot P 1985 {\it J. Stat. Phys.} {\bf 38}
823

\bibitem{J2} Jancovici B 2000 {\it J. Stat. Phys.} {\bf 100} 201

\bibitem{JKS} Jancovici B, Kalinay P, and \v{S}amaj L 2000
{\it Physica A} {\bf 279} 260

\bibitem{ST} \v{S}amaj L and Trav\v{e}nec I 2000 {\it J. Stat. Phys.} 
{\bf 101} 713

\bibitem{SJ} \v{S}amaj L and Jancovici B 2001 {\it J. Stat. Phys.} 
{\bf 103} 717

\bibitem{S} \v{S}amaj L 2001 {\it J. Stat. Phys.} {\bf103} 737

\bibitem{SJ2} \v{S}amaj L and Jancovici B 2002 {\it J. Stat. Phys.} 
{\bf106} 301

\bibitem{J3} Jancovici B 1982 {\it J. Stat. Phys.} {\bf 28} 43

\bibitem{J4} Jancovici B 1982 {\it J. Stat. Phys.} {\bf 29} 263

\bibitem{J5} Jancovici B 1995 {\it J. Stat. Phys.} {\bf 80} 445

\bibitem{J6} Jancovici B 2002 {\it Charge fluctuations in finite Coulomb
systems} (cond-mat/0201212) to be published in {\it J. Stat. Phys.} 

\bibitem{E} Erd\'elyi A 1953 {\it Higher Transcendental Functions} 
vol. II (McGraw-Hill: New York)

\bibitem{Ch} Choquard Ph, Piller B, and Rentsch R 1987  
{\it J. Stat. Phys.} {\bf 46} 599

\bibitem{FJ} Forrester P J and Jancovici B 1996 
{\it Int. J. Mod. Phys. A} {\bf 11} 941
 
\bibitem{Ch2}  Choquard Ph, Piller B, and Rentsch R 1986
{\it J. Stat. Phys.} {\bf 43} 197

\bibitem{AS} Abramowitz M and Stegun I A 1964 {\it Handbook of 
Mathematical Functions} (National Bureau of Standards: Washington)

\bibitem{Ch3} Choquard Ph, Piller B, Rentsch R, and Vieillefosse P 1989
{\it J. Stat. Phys.} {\bf 55} 1185

\end{thebibliography}
\end{document}